\documentclass[fleqn,12pt]{wlscirep}
\usepackage[utf8]{inputenc}
\usepackage[T1]{fontenc}
\usepackage{xcolor}
\usepackage[font=scriptsize]{caption}
\usepackage{mhchem}
\usepackage{pifont}
\usepackage{chemformula}
\usepackage{verbatim}

\usepackage{filecontents}
\usepackage[left]{lineno}
\usepackage[normalem]{ulem}

\newcommand{\BT}[1]{\textcolor{black}{#1}}
\newcommand{\BTbis}[1]{\textcolor{black}{#1}}

\title{A rich hydrocarbon chemistry and high C to O
  ratio in the inner disk around a very low-mass star}

\author[1*,2]{B. Tabone}
\author[3,2]{G. Bettoni}
\author[2]{E. F. van Dishoeck}
\author[4]{A. M. Arabhavi}
\author[3]{S. Grant}
\author[5]{D. Gasman}
\author[6]{Th. Henning}
\author[4]{I. Kamp}
\author[7,6,8]{M. G\"udel}
\author[9]{P.O. Lagage}
\author[10]{T. Ray}
\author[5]{B. Vandenbussche}
\author[1]{A. Abergel}
\author[11]{O. Absil}
\author[5]{I. Argyriou}
\author[12]{D. Barrado}
\author[13]{A. Boccaletti}
\author[6]{J. Bouwman}
\author[14,10]{A. Caratti o Garatti}
\author[15]{V. Geers}
\author[7]{A.M. Glauser}
\author[16]{K. Justannont}
\author[17]{F. Lahuis}
\author[4]{M. Mueller}
\author[9]{C. Nehm\'e}
\author[18]{G. Olofsson}
\author[9]{E. Pantin}
\author[6]{S. Scheithauer}
\author[5]{C. Waelkens}
\author[19,20]{L. B. F. M. Waters}
\author[16]{J. H. Black}
\author[11]{V. Christiaens}
\author[8]{R. Guadarrama}
\author[12]{M. Morales-Calder\'on}
\author[19]{H. Jang}
\author[4,23]{J. Kanwar}
\author[8]{N. Pawellek}
\author[6]{G. Perotti}
\author[21]{A. Perrin}
\author[10]{D. Rodgers-Lee}
\author[6]{M. Samland}
\author[6]{J. Schreiber}
\author[6]{K. Schwarz}
\author[22]{L. Colina}
\author[18]{G. \"Ostlin}
\author[15]{G. Wright}

\affil[1]{Universit\'e Paris-Saclay, CNRS, Institut d’Astrophysique Spatiale, 91405 Orsay, France}
\affil[*]{benoit.tabone@universite-paris-saclay.fr}
\affil[2]{Leiden Observatory, Leiden University, PO Box 9513, NL--2300 RA Leiden, The Netherlands}
\affil[3]{Max-Planck-Institut f\"ur extraterrestrische Physik (MPE), Gie{\ss}enbachstrasse 1, 85748 Garching, Germany}
\affil[4]{Kapteyn Astronomical Institute, University of Groningen, P.O. Box 800, 9700 AV Groningen, The Netherlands}
\affil[5]{Institute of Astronomy, KU Leuven, Celestijnenlaan 200D, 3001 Leuven, Belgium}
\affil[6]{Max-Planck-Institut f\"{u}r Astronomie (MPIA), K\"{o}nigstuhl 17, 69117 Heidelberg, Germany}
\affil[7]{ETH Z\"urich, Institute for Particle Physics and Astrophysics, Wolfgang-Pauli-Str. 27, 8093 Z\"urich, Switzerland}
\affil[8]{Dept. of Astrophysics, University of Vienna, T\"urkenschanzstr 17, A-1180 Vienna, Austria}
\affil[9]{Universit\'e Paris-Saclay, Universit\'e Paris Cit\'e, CEA, CNRS, AIM, F-91191 Gif-sur-Yvette, France}
\affil[10]{Dublin Institute for Advanced Studies, 31 Fitzwilliam Place, D02 XF86 Dublin, Ireland}
\affil[11]{STAR Institute, Universit\'e de Li\`ege, All\'ee du Six Ao\^ut 19c, 4000 Li\`ege, Belgium}
\affil[12]{Centro de Astrobiolog\'ia (CAB), CSIC-INTA, ESAC Campus, Camino Bajo del Castillo s/n, 28692 Villanueva de la Ca\~nada,
Madrid, Spain}
\affil[13]{LESIA, Observatoire de Paris, Universit\'e PSL, CNRS, Sorbonne Universit\'e, Universit\'e de Paris, 5 place Jules Janssen, 92195 Meudon, France}
\affil[14]{INAF – Osservatorio Astronomico di Capodimonte, Salita Moiariello 16, 80131 Napoli, Italy}
\affil[15]{UK Astronomy Technology Centre, Royal Observatory Edinburgh, Blackford Hill, Edinburgh EH9 3HJ, UK}
\affil[16]{Chalmers University of Technology, Department of Space, Earth and Environment, Onsala Space Observatory, 439 92 Onsala, Sweden}
\affil[17]{SRON Netherlands Institute for Space Research, PO Box 800, 9700 AV, Groningen, The Netherlands}
\affil[18]{Department of Astronomy, Stockholm University, AlbaNova University Center, 10691 Stockholm, Sweden}
\affil[19]{Dept. of Astrophysics/IMAPP, Radboud University, PO Box 9010, 6500 GL Nijmegen, The Netherlands}
\affil[20]{SRON Netherlands Institute for Space Research, Niels Bohrweg 4, 2333 CA Leiden, The Netherlands}
\affil[21]{Laboratoire de M\'et\'eorologie Dynamique/IPSL, CNRS, Ecole Polytechnique, Institut polytechnique de Paris, Sorbonne universit\'e, PSL research
university, F-91120 Palaiseau, France}
\affil[22]{Centro de Astrobiolog\'ia (CAB, CSIC-INTA), Carretera de Ajalvir, E-28850 Torrej\'on de Ardoz, Madrid, Spain}
\affil[23]{Space Research Institute, Austrian Academy of Sciences, Schmiedlstr. 6, A-8042, Graz, Austria}

\newpage


\begin{abstract}
  \bf {
Carbon is an essential element for life but how much can be delivered to young planets is still an open question. \BTbis{The chemical characterization of planet-forming disks is a crucial step in our understanding of the diversity and habitability of exoplanets. Very low-mass stars ($<0.2~M_{\odot}$) are interesting targets
because they host a rich population of terrestrial planets.} Here we present the JWST detection of abundant hydrocarbons in the disk of a very low-mass star obtained as part of the MIRI mid-INfrared Disk Survey (MINDS). In addition to very strong and broad emission from  C$_2$H$_2$ and its $^{13}$C$^{12}$CH$_2$ isotopologue, C$_4$H$_2$, benzene, and \BT{possibly} CH$_4$ are identified, but water, PAH and silicate
features are weak or absent. \BTbis{The lack of small silicate grains implies that we can look deep down into this disk.} These detections testify to an active warm hydrocarbon chemistry with a high C/O ratio in the inner 0.1 au of this disk, perhaps due to destruction of carbonaceous grains. The exceptionally high \BT{C$_2$H$_2$/CO$_2$ and} C$_2$H$_2$/H$_2$O \BT{column density} ratios suggest that oxygen is locked up in icy pebbles and planetesimals outside the water iceline. This, in turn, will have significant consequences for the composition of forming \BTbis{exoplanets}.}
\end{abstract}

\begin{document}

\flushbottom
\maketitle
%
%
\thispagestyle{empty}

\bigskip
\bigskip
\noindent
M dwarfs are the most common stars in the Galaxy and are known to host
exoplanets in abundance \cite{Dressing2015,Sabotta2021}. However, the
terrestrial planet-forming zones of the disks around M dwarfs have
been largely inaccessible with previous observations due to limited
spatial and spectral resolution and the dim nature of these objects.
The source 2MASS-J16053215-1933159 (hereafter denoted as J160532) is a
member of the $\sim$3--11 Myr old Upper Scorpius star forming region
at a distance of 152$\pm$1 pc \cite{GaiaDR3}, with an age of
$2.6 \pm 1.6$ Myr \cite{Miret2022}. Its spectral type of M4.75 points
to a very low-mass young star ($M_*$=0.14 M$_\odot$, $L_*$=0.04
L$_\odot$) \cite{Carpenter2014,Luhman2018,Pascucci2013} that is still
undergoing accretion at a rate of $\sim 10^{-10}-10^{-9}$ M$_{\odot}$
yr$^{-1}$. Its broadband infrared spectral energy distribution (SED)
indicates the presence of circumstellar material in the form of a
disk-like structure in which planets could originate. The non
detection of millimeter continuum emission \cite{Barenfeld2016}
suggests that the current disk mass in millimeter-sized grains is less
than 0.75 M$_{\rm Earth}$. For a standard gas/dust ratio of 100, this
would imply a gas mass that is less than 20\% of that of Jupiter.

\begin{figure}[htb!]
\centering
\includegraphics[width=\textwidth]{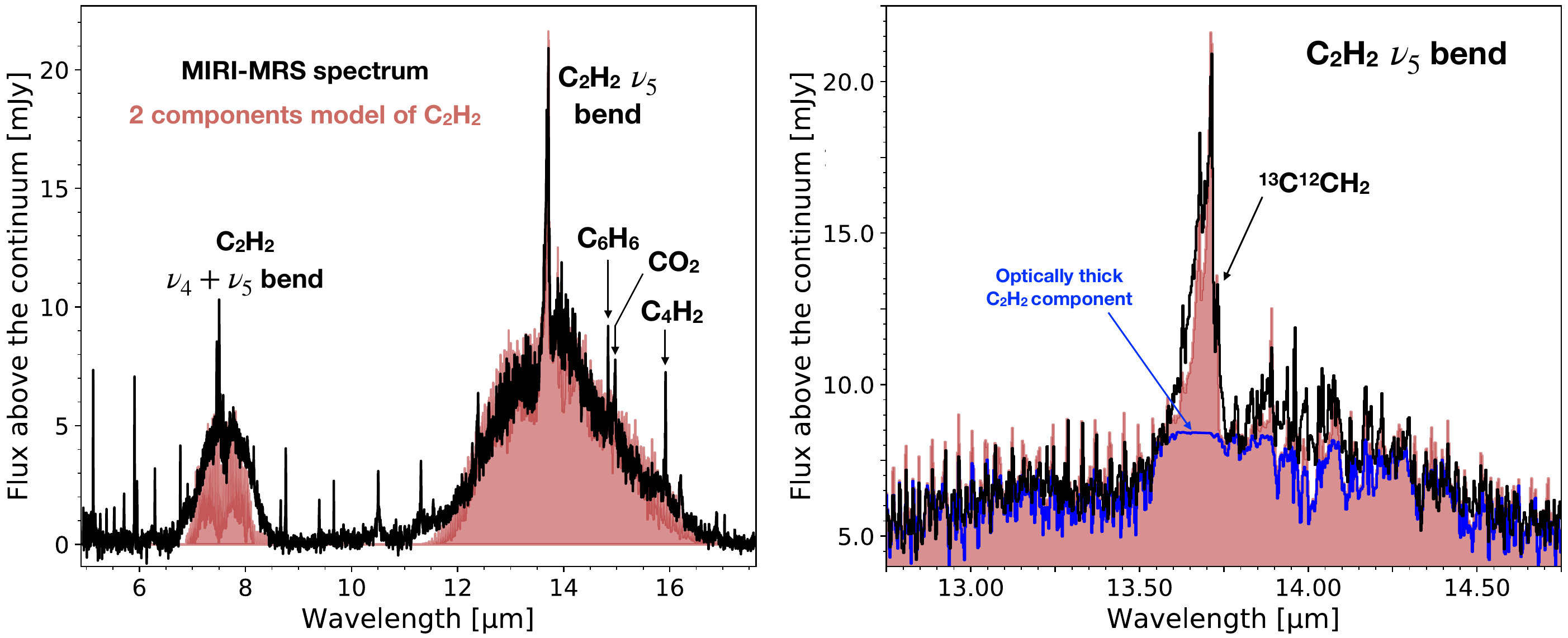}
\caption{\BTbis{JWST MIRI-MRS spectrum of J160532 showing prominent  C$_2$H$_2$ emission.} Left: Continuum subtracted MIRI-MRS spectrum of J160532
  in the 5--17.5 $\mu$m range in black (see also Extended Data Figure 5) compared with a simulated model spectrum of C$_2$H$_2$ in red. The two broad continuum bumps at 7.7 and 13.7 $\mu$m are reproduced by a high column density, highly optically thick C$_2$H$_2$ \BTbis{component I} at 525 K with an emitting area of \BTbis{$\pi (0.033 \text{au})^2$} which \BTbis{masks} the prominent $Q$-branch at 13.7 $\mu$m (see also Extended Data Figure 6). Right: zoom-in on the 13--14.5 $\mu$m range, showing that the prominent $\nu_5$ $Q$-branch of C$_2$H$_2$ is well matched by a second, more extended lower column density and less optically thick \BTbis{component II} at \BT{400 K} with an emitting area of \BT{$\pi (0.07 \text{au})^2$}. The blue line in this
  zoom-in shows the contribution of the optically thick
  component I.}
\label{Fig1}
\end{figure}

We observed J160532 with the JWST MIRI\cite{Wright2015} Medium
Resolution Spectrometer (MRS) with a spectral resolving power
$R\sim$1500--4000 covering 5--28 $\mu$m as part of the guaranteed time
MIRI mid-INfrared Disk Survey (MINDS) (see Methods section for more
details).  The continuum subtracted 5--18 $\mu$m part of the MIRI spectrum 
is presented in Figure \ref{Fig1}.
Compared with mid-infrared spectra of other disks around low-mass
stars obtained with the {\it Spitzer} Space Telescope
\cite[e.g.,]{KesslerSilacci2006,Furlan2006}, its shape is unusual
\cite{Pascucci2013}. The broad wavelength spectrum clearly shows two strong, broad bumps
centered at 7.7 $\mu$m and 13.7 $\mu$m, not seen toward any other disk so
far\cite{Dahm2009}. In contrast, no clear silicate emission features are found at 10
$\mu$m and 18 $\mu$m, nor any features due to Polycyclic Aromatic
Hydrocarbons (PAHs) at 6.2, 7.7, 8.6 or 11.3 $\mu$m (see spectrum in Extended Data Fig. 5). These broadband characteristics suggest that the
J160532 disk is settled and evolved, with silicate grains in the disk
atmosphere that must have grown to at least 5 $\mu$m.

The much higher spectral resolution of MIRI-MRS compared with previous
{\it Spitzer} data reveals numerous narrow hydrogen recombination
lines as well as molecular features on top of the continuum
(Figure~\ref{Fig1}). C$_2$H$_2$ emission at 13.7 $\mu$m is
particularly strong, consistent with earlier findings that this
molecule is enhanced in disks around brown dwarfs and very low-mass
stars \cite{Pascucci2009,Dahm2009,Pascucci2013}. We focus here on the analysis
of the molecular lines and demonstrate that they can be ascribed to a
mix of small aliphatic and aromatic hydrocarbon molecules plus
CO$_2$ but that any water lines are weak.  Column density ratios are
found to be very different from those found in disks around the more
massive T Tauri stars.

\section*{Results}
\label{sec:results}

Molecular species were identified by matching the most prominent
features in the continuum subtracted JWST-MIRI spectrum with synthetic
spectra (see Methods section for more details). Most of the bands
involve vibration-rotation transitions. The synthetic spectrum of each
molecule is calculated from a plane-parallel slab model, where the gas
is assumed to have a uniform temperature $T$ and the excitation of the
molecules to be in Local Thermodynamic Equilibrium (LTE) at a single
excitation temperature $T_{\rm ex}$ equal to $T$
\cite{Carr2011,Salyk2011a} . The other fitting parameters
are the line of sight column density $N$ within a projected emitting
area $\pi R^2$ given by its radius $R$. Note that $R$ does not need to
correspond to a disk radius, but could also represent a ring with the
same area. The best fitting parameters are summarized in Table~1.

The shape and position of any $Q-$branch, where lines with zero change
in rotational quantum number $J$ pile up, are particularly sensitive to
temperature. The full ro-vibrational bands of all considered species but H$_2$O require a treatment of line overlap in the optically thick case. For most species with only a single feature, there is
often a degeneracy between a high $T$, low $N$ optically thin solution
and a lower $T$, high $N$ optically thick case. Uncertainties and
degeneracies associated to the fits are evaluated using a $\chi^2$
approach following earlier studies \cite{Carr2011,Salyk2011a} \BTbis{(see Extended Data Figure 7)}.

\smallskip
\noindent
-- {\bf C$_2$H$_2$ and $^{13}$C$^{12}$CH$_2$}: The $Q$-branch of
C$_2$H$_2$ at 13.7 $\mu$m associated with the $\nu_5$ bending mode on
top of the broad continuum is the most prominent feature in the entire
MIRI-MRS spectrum (Figure~\ref{Fig1}). At $R\sim$3000, MIRI-MRS also
reveals a series of $P-$ and $R-$branches on top of both the 13.7
$\mu$m and 7.7 $\mu$m broad bumps. The fact that these bumps coincide
in location with emission from gaseous C$_2$H$_2$ suggests that the
carrier may be due to hot and very abundant C$_2$H$_2$ itself. No
solid material can be identified that coincides with these broad
bumps, and the spacing between the features is too broad to be due to
silicate absorption in a near edge-on system suggested based on {\it
  Spitzer} data \cite{Pascucci2013}. We demonstrate here that both the
broad and narrow components are well reproduced by a two-component
model consisting of highly optically thick and more optically thin
C$_2$H$_2$ emission.

Figure~\ref{Fig1} shows that the overall shape of the 13.7 $\mu$m
continuum bump can be well fit by a slab of gas at $T = 525$ K with a
column density of \BT{$N$(C$_2$H$_2$)=$2.4 \times 10^{20}$ cm$^{-2}$} within an emitting area of $\pi(0.033~\rm{au})^2$ (i.e, $R=0.033$ au). Our fit includes the contribution of $^{13}$C$^{12}$CH$_2$, assuming a C$_2$H$_2$/$^{13}$C$^{12}$CH$_2$ ratio of 35 \cite{Woods2009}. \BTbis{In the following, this highly optically thick and compact component is called component I}. Such an exceptionally high column density of C$_2$H$_2$ is required to fully saturate the blended
molecular lines and produce a pseudo-continuum that masks any prominent features like the $Q$-branch (see Extended Data Figure 6). Some fraction of the 7.7 $\mu$m combination $\nu_4 + \nu_5$ band is also recovered, but with a too high contrast between the amplitude of the narrow features and the level of the pseudo-continuum. At such high column densities, hot bands of $^{13}$C$^{12}$CH$_2$ \BT{that are not included in spectroscopic databases such as HITRAN} should contribute significantly as well to the 7.7 $\mu$m bump and result in a blending of the individual lines. Proper modeling must await more complete $^{13}$C$^{12}$CH$_2$ molecular spectroscopy including highly excited bands.

The presence of a prominent $Q-$branch at 13.7 $\mu$m indicates a
second physical component, \BTbis{called component II}, producing less optically thick C$_2$H$_2$
emission. Our MIRI-MRS data allow to distinguish also the shortward
peaks at 13.63 $\mu$m and 13.68 $\mu$m due to hot bands that were
blended with the main peak in lower resolution {\it Spitzer} spectra
\cite{Pascucci2009,Pascucci2013} (see Figure~\ref{Fig1}, right). These features
are not tracing the bulk reservoir of C$_2$H$_2$ but can either unveil
the hotter layer at the surface of the thick component or a more radially extended
emission. As an illustration, these features are indeed well
reproduced by a somewhat more extended lower column density of
$N$(C$_2$H$_2$)=$2.5 \times 10^{17}$cm$^{-2}$ with $R=0.07$~au at a temperature of 400 K. This component would then trace a physically distinct region, at the outer boundary of the C$_2$H$_2$-rich region of the disk. This is the assumption that we make for the analysis of most of the other molecules, which allows us to subtract the contribution of the two C$_2$H$_2$ bumps in the spectra \BTbis{(see Methods section and Extended Data Figure 5) and fit their features without taking into account the  masking of the features by optically thick C$_2$H$_2$ lines from component I.} 

The $Q-$branch of $^{13}$C$^{12}$CH$_2$ is also detected at 13.73
$\mu$m above the optically thick \BTbis{component I}. If this emission would
originate from the optically thinner \BTbis{component II}, a
C$_2$H$_2$/$^{13}$C$^{12}$CH$_2$ abundance ratio of about 3 would
be required to match the peak intensity, a value that is an order of magnitude lower than the interstellar $^{12}$C/$^{13}$C
ratio. More likely, this indicates a more complex layered structure than the slab model can simulate. Sophisticated models including radial and
vertical gradients of temperature and a more complete spectroscopy of
$^{13}$C$^{12}$CH$_2$ are needed to consistently interpret the
prominent peaks of C$_2$H$_2$ and $^{13}$C$^{12}$CH$_2$ .



\begin{figure}[htb!]
\centering
\includegraphics[width=\textwidth]{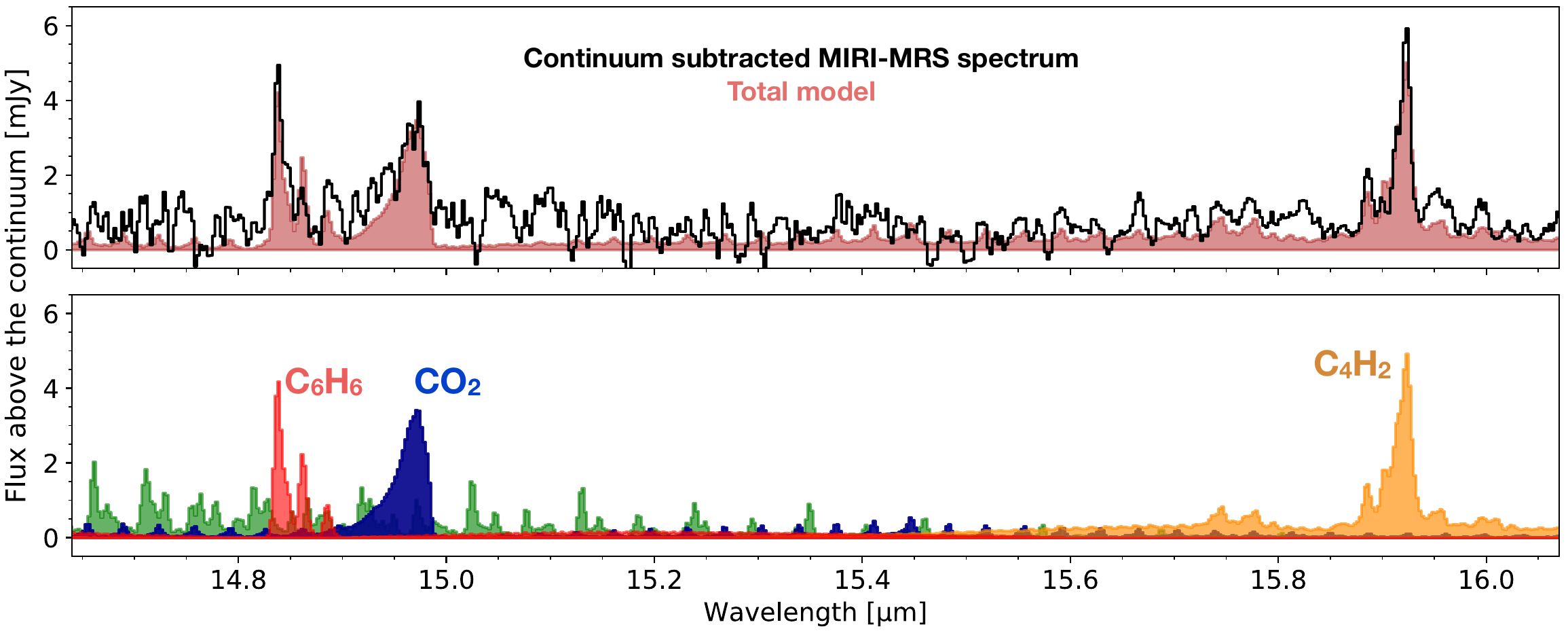}
\caption{\BTbis{Detection of C$_6$H$_6$ (benzene), CO$_2$, and C$_4$H$_2$.} Continuum subtracted MIRI spectrum, showing a zoom-in on a number of key molecular transitions in the 14.7-16 $\mu$m region (top panel) and their best fit slab models assuming an emitting area of $\pi (0.07 \text{au})^2$ (bottom panel). The contribution of the two C$_2$H$_2$ bumps \BTbis{(component I)} has been subtracted (see Extended Data Figure 5). However, narrow C$_2$H$_2$ features are still present shortward of $\simeq 15.5~\mu$m as exemplified by a slab model of C$_2$H$_2$ (green spectrum). \BTbis{We note that CO$_2$ emission likely originates from the C$_2$H$_2$ thick component I with a smaller emitting area of $\pi (0.033 \text{au})^2$ (see alternative fit in Table 1).}}
\label{Fig2}
\end{figure}

\smallskip
\noindent
-- {\bf C$_4$H$_2$}: The emission features at 15.92 $\mu$m highlighted
in Figure~\ref{Fig2} correspond to the $Q-$branch of the fundamental
bending mode $v_8$ of di-acetylene, C$_4$H$_2$. As for other
molecules, the shape of this feature depends strongly on temperature,
becoming broader at higher $T$. Interestingly, an origin within a small emitting area of $R \lesssim 0.07~$au is excluded as line overlap would \BTbis{mask} the prominent features. This supports the scenario that the C$_4$H$_2$ features originate from the optically thinner \BTbis{component II}. This also motivates the choice of our reference emitting area for the thin component to $R=0.07~$au. A model with $T=330$~K is able to
reproduce the feature at 15.92 $\mu$m as well as that at 15.88 $\mu$m. Moreover, the two emission features between 15.7--15.8
$\mu$m are well reproduced. 

\smallskip
\noindent
-- {\bf C$_{6}$H$_{6}$}:
We identified three features around 14.85 $\mu$m to be the
$Q-$branches of the fundamental and hot bending mode $v_4$ of benzene,
C$_{6}$H$_{6}$, presented in Figure~\ref{Fig2}. Their relative intensity is sensitive to temperature and indicates $T \simeq 400~$K. \BT{As for C$_4$H$_2$, compact emission ($R \lesssim 0.05~$au) is excluded.}

\smallskip
\noindent
-- {\bf CH$_{4}$}: Extended Data Figure  8 shows possible indications of
CH$_4$ emission. CH$_4$ was previously seen in the GV Tau N disk
\cite{Gibb2013}, but only in absorption. We observe emission lines at
7.65--7.67 $\mu$m that are aligned with the $Q$ branch of the $v_4$
mode of CH$_4$.  C$_{2}$H$_{2}$ also has many emission lines in this
region, but cannot reproduce by itself this broad feature.

\smallskip
\noindent
-- {\bf HCN}: The ro-vibrational band from the fundamental $v_2$ bending
mode of HCN is severely blended with the strong emission lines of
C$_{2}$H$_{2}$. Extended Data Figure 9 shows the maximum amount of HCN that could be present in the optically thinner C$_2$H$_2$ component II. If present in the C$_2$H$_2$ thick component I, HCN emission features would be severely \BTbis{masked} and its column density in that region cannot be robustly constrained.

\smallskip
\noindent
-- {\bf CO$_{2}$}: Figure~\ref{Fig2} includes the fit to the CO$_2$
bending mode at 14.98 $\mu$m that is clearly detected. Assuming that CO$_2$ emission originates from the C$_2$H$_2$ thin component II, the shape of its $Q-$branch indicates a high temperature around 650 K. \BT{Interestingly, the $\chi^2$ fit points also toward a smaller emitting area very close to that for the optically thick C$_2$H$_2$ component I with similar temperature and a column density as high as 2$\times$10$^{18}$ cm$^{-2}$. In the latter case, C$_2$H$_2$ can partially \BTbis{mask} CO$_2$ emission but we checked that the fitted column density is then underestimated by less than a factor of 2. The $Q$-branch of $^{13}$CO$_2$ at $15.4~\mu$m is not detected, in line with the column densities reported in Table 1.}

\smallskip
\noindent
-- {\bf H$_2$O}: Extended Data Figure  10 presents blow-ups of the
regions of the MIRI spectrum where water can be observed: at 6 $\mu$m
through the $\nu_2$ ro-vibrational lines, and at $>$10 $\mu$m through
highly excited pure rotational transitions
\cite{Carr2008,Salyk2011a}. In the J160532 spectrum, neither set of
lines are clearly seen but there are a few weak features around 17 $\mu$m
that could potentially be consistent with water emission. \BT{In those spectral regions, C$_2$H$_2$ cannot mask H$_2$O emission.} The
values listed in Table~1 could also be viewed as an upper limit on the
amount of water hidden in this spectrum. No OH lines are found.

\smallskip
\noindent
- {\bf H$_2$}: Several H$_2$ pure rotational lines are clearly seen in
the MIRI-MRS spectrum, which will be analyzed in detail elsewhere. 
For a temperature of $\sim$550 K, indicated by the $S$(1)/$S$(3) line ratio, the total mass of  warm H$_2$ is about $3\times 10^{-5}$ M$_{\rm Jup}$.

\smallskip
\noindent
-- {\bf Other species}: \BT{At the shortest MIRI wavelength range, between 4.9 and 5.1 $\mu$m, several CO $v=1-0$ $P$-branch lines are found, indicative of high temperature gas ($T > 1000~$K) that will be analyzed elsewhere.} Several other hydrocarbon species (C$_2$H$_4$,
HC$_3$N) were searched for in the J160532 spectrum, but not
identified. Also, NH$_3$, whose $\nu_6$ mode at 8.8 $\mu$m can be
observed with MIRI, was not found in the current spectrum. \BT{The [Ne II] line at 12.8 $\mu$m is not detected.}




\begin{table}[h]
  \begin{center}
  \caption{Best fit slab model results for molecules in the J160532 disk.}
\begin{tabular}{|c|cc|cc|}
  \toprule
  
   & \multicolumn{2}{c}{Component I$^a$} & \multicolumn{2}{|c|}{Component II}\\
  Molecule       &  $T$    &           $N$         &  $T$ &  $N$                            \\[0.5ex]
                 &  (K)    & (10$^{17}$ cm$^{-2}$) &  (K) & (10$^{17}$ cm$^{-2}$)           \\[0.5ex]
  \midrule
  C$_2$H$_2$     &  525     &   $2400^{+3200}_{-1400}$      & 400     & 2.5             \\[0.5ex]
  C$_{4}$H$_{2}$ &  -       &   -                           & 330     & 0.7             \\[0.5ex]
  C$_6$H$_6$     &  -       &   -                           & 400     & 0.7             \\[0.5ex]
  CH$_{4}$       &  -       &   -                           & 400$^b$ & 1.5             \\[0.5ex]
  CO$_2$         & 430      & 20$^{+55}_{-18}$              & 650     & 0.36            \\[0.5ex]
  HCN            &  -       &  -                            & 400$^b$ &  $\leq$ 1.5       \\[0.5ex]
  H$_{2}$O       & 525$^b$  &    $\leq 30$                  & 400$^b$ & $\leq \BTbis{8}$        \\[0.5ex]
  \bottomrule
\end{tabular}
\end{center}
{{\BTbis{Note a: for H$_2$O and CO$_2$, the reported values correspond to an alternative fit to component II assuming that all the emission originates from the C$_2$H$_2$ thick component I ($R=0.033~\text{au}$).}} \\
{Note b: fit performed by fixing the temperature \BT{to that of the corresponding} C$_2$H$_2$ component.}} \\
{Uncertainties on the column densities correspond to the $1 \sigma$ confidence interval obtained by fixing the emitting area and are only valid in the framework of our simple slab modelling. For clarity, uncertainties smaller than 0.5 dex are not reported.}
\label{Tab1}
\end{table}



\section*{Discussion}
\label{sec:discussion}

The most striking feature of the J160532 MIRI spectrum is the
dominance of hydrocarbon emission, most notably C$_2$H$_2$, but also
C$_4$H$_2$, C$_6$H$_6$ (benzene), and possibly CH$_4$. In contrast, at best weak
H$_2$O emission is found, and CO$_2$ has a similar column density as
most hydrocarbons \BTbis{in component II} (Table~1).  
Hydrocarbon molecules like
C$_4$H$_2$ and benzene have been found previously in some
astrophysical environments, including asymptotic giant branch stars,
comets and moons in our own Solar System \cite{Cernicharo2001a,
  Coustensis2007, Schuhmann2019}, but not yet in the planet-forming
zones of disks. These detections therefore highlight that the
inner disks around very low-mass objects are indeed very rich in
carbon-bearing molecules \BT{as suggested based on \textit{Spitzer} data} \cite{Dahm2009,Pascucci2009,Pascucci2013}. The high column densities point toward being able to probe deep down layers, likely due to a lack of grains in the inner disk.  

\begin{figure}[htb!]
\centering
\includegraphics[width=0.5\textwidth]{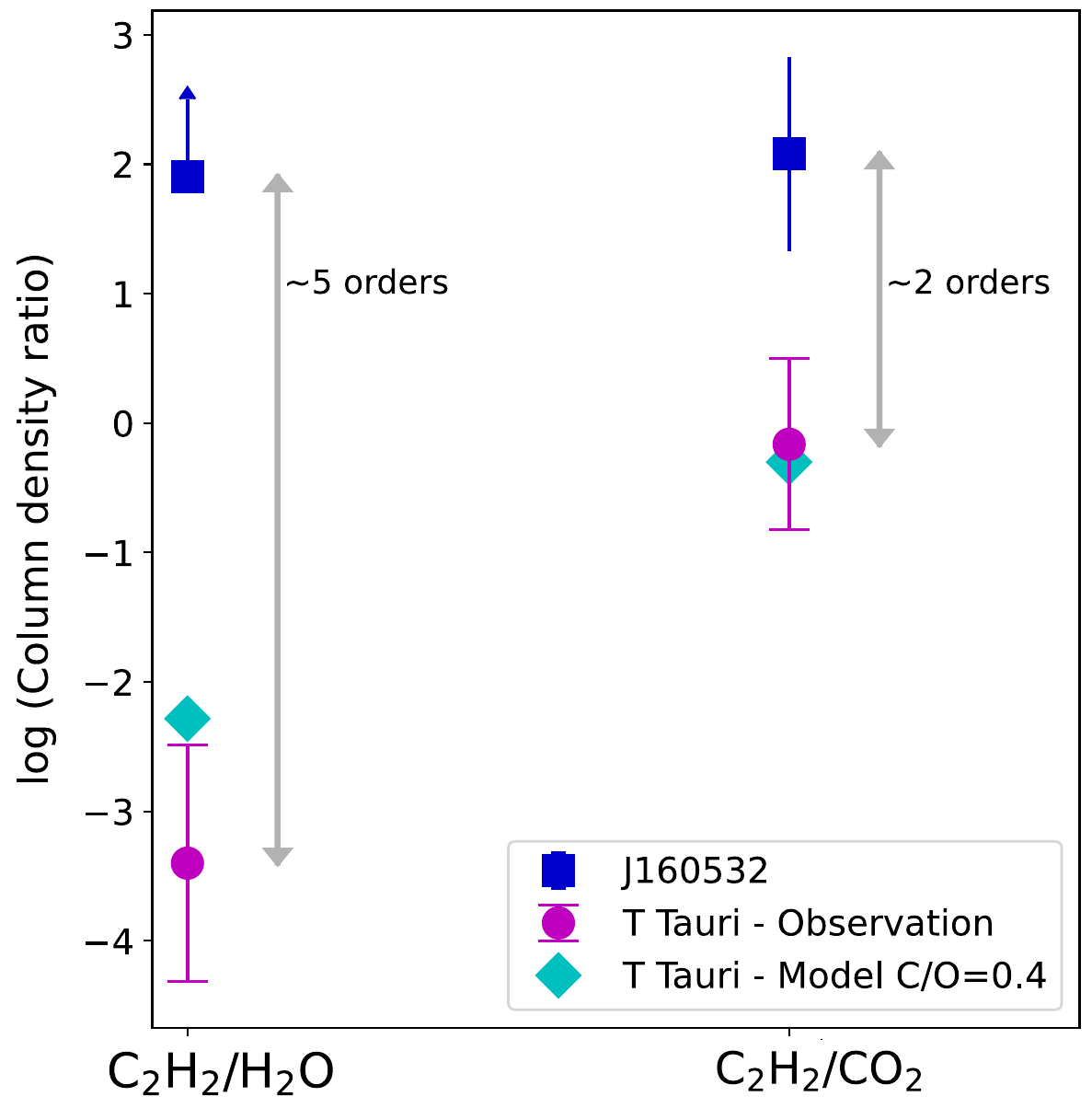}
\caption{\BTbis{Column density ratios in the inner disk of J160532.} Comparison of the observed column density ratios in component I with an emitting area of $\pi (0.033~\rm{au})^2$ in the disk of J160532 with those found for T Tauri disks with detected C$_2$H$_2$ and H$_2$O \cite{Salyk2011a}. The ratios are also compared with those found in
  thermo-chemical disk models \BT{that assume solar C/O elemental ratios} \cite{Woods2009}. Note that column
  density ratios should not be equated with abundance ratios since
  abundances of individual molecules vary strongly radially and
  vertically. \BT{The error bars on the measured J160532 ratios are $1 \sigma$ confidence interval estimated from our $\chi^2$ fit assuming similar emitting areas for CO$_2$ and C$_2$H$_2$. The error bars on the ratios measured in the T Tauri sample correspond to the $\pm 1 \sigma$ of the distribution.}}
\label{Fig3}
\end{figure}

Figure~\ref{Fig3} summarizes the observed column density ratios of key
molecules in J160532 with those found in disks around more massive T
Tauri stars \cite{Salyk2011a}. For the latter, the sources listed in
Table 8 of Salyk et al. \cite{Salyk2011a} with detected C$_2$H$_2$ and
H$_2$O are taken. The C$_2$H$_2$/H$_2$O ratio for the optically thick
component I is up to 5 orders of magnitude higher in the J160532 disk
than for T Tauri disks, and even more if H$_2$O is treated as an
upper limit. Similarly, the
C$_2$H$_2$/CO$_2$ ratio is \BT{2} orders of magnitude higher. These
ratios are much higher than just the flux ratios shown in
Extended Data Figure 11 since our analysis, including the hot
bands and a detailed treatment of line overlap, clearly demonstrates that the C$_2$H$_2$
emission is highly optically thick boosting its column density by
orders of magnitude. Note that these column density ratios should not
be viewed as local abundance ratios since each molecular band may
originate from a different part of the disk, with abundances known to
vary radially and vertically \cite{Woitke2018}.

Nevertheless, the J160532 disk is clearly rich in hydrocarbon
molecules and the observed chemical differences indicate that
hydrocarbon molecules either form more efficiently in disks around
very low-mass stars, or that the conditions for their survival are
more favorable there. One difference is the UV spectrum of the central
star, which has many fewer high energy photons that can
photodissociate molecules for an M-type star than for an early K- or
G-type star. However, J160532 still does have some accretion
consistent with its relatively young age of 2.5 Myr; its estimated FUV
luminosity of log($L_{\rm FUV}$)=--3.59 L$_\odot$ \cite{Pascucci2013}
is comparable to that of T Tauri stars. As an M star, J160532 may also
have chromospheric activity and flares producing enhanced UV and
X-rays, the latter estimated \cite{Pascucci2013} at
log($L_{\rm X}$)=28.8 erg s$^{-1}$.

The alternative possibility is enhanced C$_2$H$_2$ production. Carbonaceous grains and PAHs can be destroyed in the inner disk due to UV
radiation, chemical processes or sublimation producing abundant C$_2$H$_2$
\cite{Kress2010,Anderson2017}. Interestingly, the so-called hydrocarbon
``soot'' line as defined by the sublimation front of refractory carbon is estimated to lie at 500 K \cite{Li2021}, the
temperature found for the abundant optically thick C$_2$H$_2$
component I. The exact location of the
``soot'' line is however uncertain and depends on the type of
carbonaceous material, with some laboratory data putting sublimation of
amorphous carbon grains at higher temperatures, up to 1200 K
\cite{Gail2017}. One possibility is therefore that we are witnessing carbon grain destruction in the inner disk. \BT{Is carbon grain destruction observable in other disks? This remains an open question to be tackled with the upcoming JWST data. If carbon grains are destroyed by UV photolysis, it could be dramatically enhanced in J160532 due to dust growth and settling which increase the penetration depth of the UV. Alternatively, carbon grain destruction could be happening primarily in the high temperature midplane that is uniquely visible in the J160532 disk but hidden from our view in most disks due to the presence of dust with high infrared optical depth.}

Further insight into the hydrocarbon chemistry can be obtained from
comparison of our observed column density ratios with those found in
thermo-chemical models \cite{Walsh2015} of disks around low-mass stars
(see Methods section for more details). Interestingly, both benzene
and C$_4$H$_2$ were predicted to be abundant in inner disk regions
\cite{Woods2007,Woods2009}. \BT{For the optically thin C$_2$H$_2$ component II in which these molecules are not masked by optically thick C$_2$H$_2$, we find relatively good agreement with the models even though benzene is underestimated by the models (see Extended Data Table 2).} Under these conditions -- warm dense gas with high C$_2$H$_2$ abundance -- one would expect also efficient PAH formation up to temperatures where erosion starts to take over \cite{Frenklach1989,Morgan1991}. To what
extent the absence of PAH features in the J160532 spectrum also
implies absence of PAHs, or whether there is simply not enough UV
radiation to excite and make them visible \cite{Geers2006}, needs to
be quantified.  A detailed comparison using a physical-chemical model
appropriate for the J160532 disk is postponed to a future study.

An additional source of C$_2$H$_2$ production due to carbon grain destruction in the
J160532 disk would also be consistent with the fact that the observed
C$_2$H$_2$/H$_2$O ratio is four orders of magnitude higher than what
is found in those models. CO$_2$ is underabundant as well by
two orders of magnitude. This conclusion holds irrespective of
the stellar mass and adopted UV field in the models \cite{Walsh2015}.

Najita et al. \cite{Najita2011,Najita2013} suggested that the range in
observed HCN/H$_2$O ratios for T Tauri stars implies different C/O
ratios in the inner disk. High HCN/H$_2$O ratios would indicate that
H$_2$O is locked up in non-migrating pebbles and planetesimals in the
outer disk beyond the water iceline.  \BT{In the J160532 disk, the HCN/H$_2$O ratio cannot be robustly constrained but the C$_2$H$_2$/H$_2$O ratio is much higher than
for T Tauri disks.} \cite{vanDishoeck2021}.
\BT{In fact, thermochemical disk models of cool M-type stars predict a low H$_2$O abundance due to the lower temperature of their disks driving much of the oxygen into O$_2$\cite{Walsh2015}. Still, for solar C/O ratios, CO$_2$ is predicted to be one of the main oxygen carriers after CO and O$_2$, in stark contrast with our estimates of the CO$_2$/C$_2$H$_2$ ratio.} Such high abundance ratios as found here can therefore only be reproduced
if the C/O ratio of the gas in the inner disk is significantly
increased compared with standard values of C/O=0.4
\cite{Pascucci2013}. \BT{In fact, the models of Najita et al.\cite{Najita2011}, Woitke et al.\cite{Woitke2018} and Anderson et al. \cite{Anderson2021} show that values of C/O$\gtrsim$1 are  needed to
ensure that the bulk of the volatile oxygen is contained in CO leaving little room for H$_2$O and CO$_2$ production, and permitting the formation of abundant C$_2$H$_2$.}

\begin{figure}[htb!]
\centering
\includegraphics[width=0.6\textwidth]{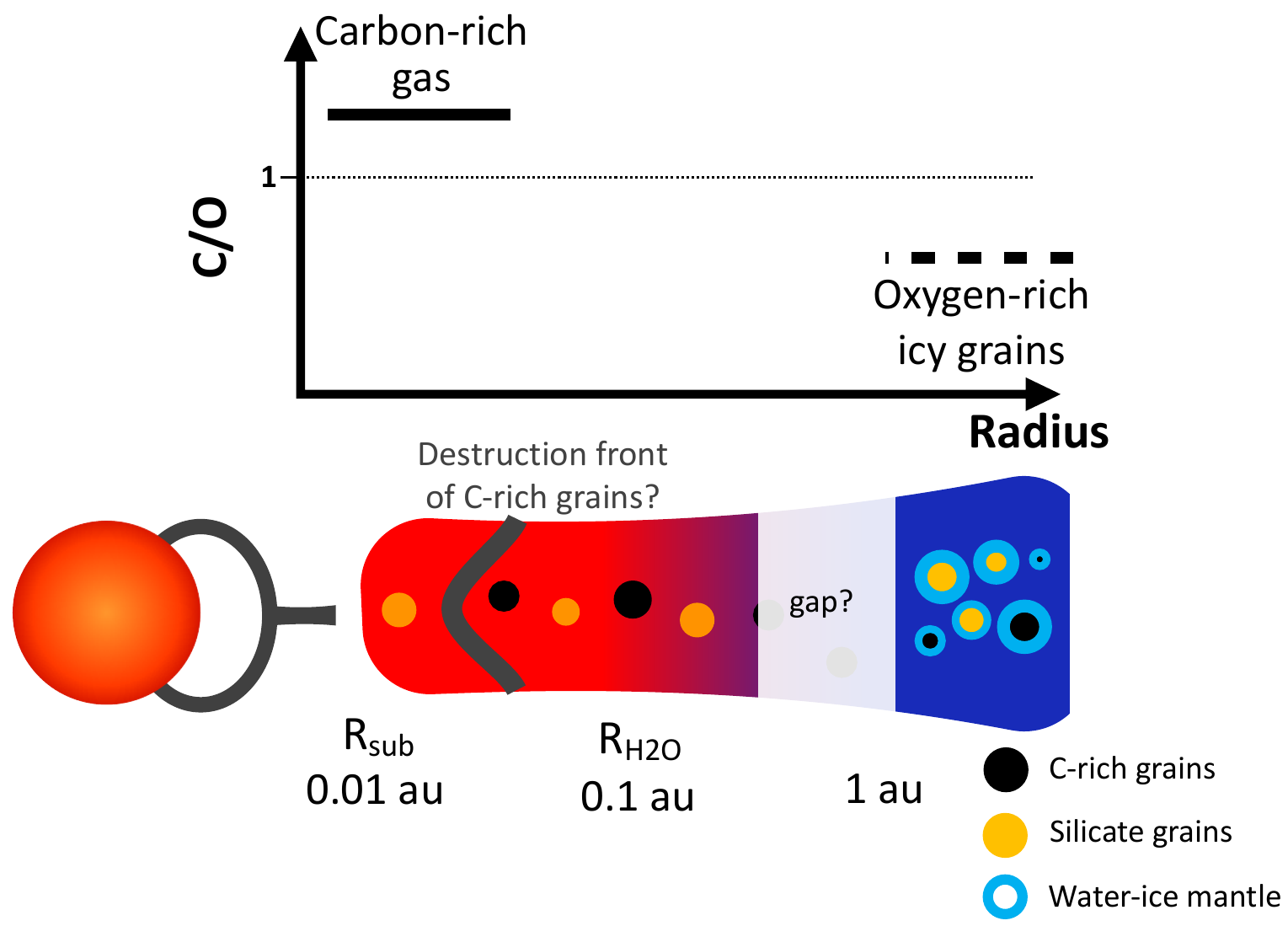}
\caption{\BTbis{The possible structure of the inner part of
  the J160532 disk.  This illustration shows} the silicate dust sublimation radius
  around 0.01 au, the destruction front of carbonaceous grains around 0.033 au (either by sublimation, UV photolysis, or chemical processes), and the water
  snowline around 0.1 au. \BTbis{In this schematic view, the outer radius of the optically thick C$_2$H$_2$ component I is put at $0.033$~au derived from the fitted emitting area of $\pi (0.033~\text{au})^2$. The location of the silicate and water sublimation fronts is estimated from the luminosity of the star.} The top shows the high C/O ratio in the gas in the very inner disk and the low C/O ratio in icy grains at larger radius. \BTbis{The inner disk contains only large ($> 5~\mu$m) silicate grains.} The location of any dust
  trap locking up water ice is unconstrained, except that it must be
  outside the water snowline at $\sim$0.1 au. Note the log scale for
  the distance to the star.}
\label{Fig4}
\end{figure}

What could cause the inner disk to be depleted in oxygen compared to carbon? Destruction of carbon-rich grains helps to boost carbon and thus the gas-phase C/O ratio by about a factor of 2 \BTbis{compared to the volatile carbon in the interstellar medium (about 50$\%$ of the total elemental carbon), enough to put the C/O ratio close to unity. However, the very high C/O ratio inferred for the inner disk of J160532 may also point toward a depletion of oxygen.} To deplete the inner disk of oxygen the most plausible
mechanism  would be to lock most of it up in water ice in pebbles and planetesimals in
the outer disk, beyond the water snowline, which for such a low mass
star is around 0.1 au \cite{Mulders2015b}. Gas with high C/O, as
  often found in the outer disk \cite{Bosman2021MAPS}, could still be
  able to cross the gap. Little is known about the
small scale structure of disks around very low mass stars like
J160532, but our findings would imply that such dust disks are not
smooth but must have significant substructure with dust and ice traps
on (sub)au scales \cite{Pinilla2013,Kurtovic2021}, as illustrated in
Figure~\ref{Fig4}. To avoid much oxygen crossing these traps, they
must develop early in the disk's evolution, perhaps due to a companion
that has formed there. ALMA observations of the outer disk, and analysis of H$_2$ and ro-vibrational CO emission from the inner disk, \BTbis{combined with detailed modelling,} are required to constrain the gas-phase C/O and O/H elemental ratios in the inner versus outer disk and further quantify the destruction of carbon grains and the efficiency of oxygen trapping in the outer disk. 

What are the implications for any planets forming around J160532? The
current amount of solid material contained in millimeter-sized grains
in its disk after 2.5 Myr of evolution is less than an Earth mass
\cite{Barenfeld2016}, but the system could have started forming
planetesimals and building planets earlier in its lifetime. In fact,
there is ample evidence for efficient \BT{terrestrial} planet formation around very
low-mass objects. The occurrence rate per M-dwarf is 2.5$\pm$0.2
planets with radii of 1--4 $R_{\rm Earth}$ and periods $<$200 days
\cite{Dressing2015} .
Surveys also find that short-period ($<$10 days) planets around stars
with $M_* <$ 0.34 M$_\odot$ are significantly overabundant relative to
more massive stars \cite{Sabotta2021}. Our JWST data reveal that the
chemistry in disks around such very low-mass stars may have an even
higher gaseous C/O ratio in the planet-forming zones than thought
before. This, in turn, could significantly affect the composition of
planets that may form around them.

To what extent the planets inside 0.1 au are also carbon rich will
depend on whether they are formed mostly from ``dry'' planetesimals in
this inner disk, or
whether their bulk composition and atmospheres result primarily from
impacting icy planetesimals from the outer disk, as found in scenarios
for making terrestrial planets \cite{Morbidelli2012,Ormel2017}. In
fact, it is not clear whether \BT
{planetesimals formed in the inner disk} are carbon
rich: our data \BT{indicate} the presence of some warm silicate refractory dust
producing continuum mid-infrared emission, \BT{whereas} a significant fraction
of the carbon may be in the gas-phase and even refractory carbon
grains may have been destroyed. This carbon-rich gas could be lost from
the system over time, with only a small fraction of carbon eventually
included in planets \cite{Lee2010}. Such a scenario likely holds for
our own Earth which is known to be very carbon poor
\cite{Li2021}. These two competing scenarios --- \BT{making} carbon-rich versus
carbon-poor \BT{terrestrial planets}--- can be tested by comparing high sensitivity JWST
observations of the chemical composition of significant samples of
disks around very low-mass young stars with that of the atmospheres of
terrestrial-sized planets around mature M stars \BTbis{like the Trappist I system\cite{Greene2023}} and brown dwarfs.





\section*{Methods}
\label{methods}

\textbf{Observations and data reduction}
\label{sec:datareduction}

J160532 was observed with the JWST Mid-InfraRed Instrument (MIRI)
\cite{Rieke2015,Wright2015,Labiano2021} in Medium Resolution
Spectroscopy mode (MRS) \cite{Wells2015} on 2022 August 1, from
05:50:01 UT, for a total observation time of 2.22 hours. The
observation is number 47 of the Cycle 1 Guaranteed Time Observation
(GTO) program 1282 (PI: Thomas Henning). Following target acquisition,
the three grating settings Short (A), Medium (B) and Long (C) were
used in each of the four channels observations carried out in
parallel, providing full coverage of the MIRI spectral window
$4.9-28.1$ $\mu$m. For each of the three sub-bands, the FASTR1 readout
pattern was adopted with a point source 4-point dither pattern, an
exposure time of 308 s and an integration time of 74.9 s.


For the reduction of the uncalibrated raw data, we used the version
1.8.4 of the JWST Science Calibration Pipeline \cite{Bushouse2022} and
the CRDS context \texttt{jwst\_1017.pmap}. The uncal files were first
processed with the default class of the pipeline
\texttt{Detector1}. Next, we performed a background correction by
subtracting each dither pattern from the other associated pattern (1
from 4, 2 from 3 and vice-versa). We then applied the default class
\texttt{Spec2}, skipping the residual fringe correction.  The data
were then processed by the \texttt{Spec3} class, which combines the
calibrated data from the different dither observations into a final
level3 spectral cube. We skipped the \texttt{outlier\_detection} and
\texttt{master\_background} methods of the class, since we already
performed the background subtraction after \texttt{Detector1}. We set
\texttt{Spec3} to produce one spectral cube for each sub-band, from
which we extracted the spectrum using the pipeline method
\texttt{extract1d}. This was done in order to provide the best input for a residual fringe correction that we applied at the spectrum level.

\medskip

\noindent \textbf{Local continuum fit} 

Extended Data Figure 5 shows
the baseline fit used to produce the
spectrum in Figures 1 and 2. Starting from the original spectrum, first a
low-order continuum due to warm dust emission has been removed over
the entire 5-20 $\mu$m range, producing Figure 1. \BTbis{Subsequently, the
two broad bumps at 7.7 $\mu$m and 13.7 $\mu$m have been subtracted in Figures 2, 8, and 10 to further analyse all the molecular features except that of
the very optically thick bumps of C$_2$H$_2$ (component I).}

\medskip

\noindent \textbf{Slab model fits}

The molecular lines are analysed using a slab approach that takes
into account optical depth effects. The level populations are assumed
to be in local thermodynamical equilibrium (LTE) and the line profile
function to be Gaussian with an intrinsic broadening of $\sigma = 2~$km
s$^{-1}$ (FWHM of $\Delta V = 4.7~$km~s$^{-1}$) to include the effect of turbulence. We note that for optically thin lines, the inferred column densities are independent of the value of $\Delta V $ whereas for optically thick emission, the inferred column densities scale approximatively as $1/\Delta V$. The line emission is assumed to originate from a layer of
gas with a temperature $T$ and a line of sight column density of
$N$. \BT{Because most of the species analysed here produce lines that are close to each other in frequency, we adopt a detailed
treatment of line overlap by first computing the wavelength dependent
opacity over a fine grid of wavelength ($\lambda / \Delta \lambda \simeq 10^6$)
\begin{equation}
\tau(\lambda)  = \sum_{i} \tau_{0,i}  e^{-(\lambda-\lambda_{0,i})^2/2 \sigma_{\lambda}^2},
\end{equation}
where $i$ is the line index, $\lambda_{0,i}$ is the rest wavelength of line $i$, $\sigma_{\lambda}$ is the intrinsic broadening of the line in $\mu$m, and $\tau_{0,i}$ is the optical depth at the center of line $i$ given by
\begin{equation}
\tau_{0}  = \sqrt{\frac{\ln 2}{\pi}} \frac{A_{ul} N \lambda_0^3}{ 4 \pi  \Delta V} (x_{l} \frac{g_u}{g_l} - x_{u}).
\end{equation}
In this equation, $x_{u}$ and $x_{l}$ denote the population level of
the upper and lower states, $g_u$ and $g_l$ their respective
statistical weights, and $A_{ul}$ the spontaneous downward rate of the transition. The flux density $F(\lambda)$  is then computed assuming an emitting area of $\pi R^2$ and a distance to the source $d$ as:
\begin{equation}
F(\lambda) = \pi \left( \frac{R}{d} \right)^2 B_{\nu}(T) (1-e^{-\tau(\lambda)}),
\end{equation}
and convolved at MIRI-MRS spectral resolution. This special treatment, though computationally expensive, is particularly crucial at high column densities for which overlapping lines can form an effectively optically thick continuum across a relatively broad spectral range (see Extended Data Figure 6). \BTbis{Our code has been benchmarked against the publicly available code \texttt{slabspec}\cite{Salyk2020}.}}
\\

The molecular data, i.e., line positions, Einstein $A$
coefficients, statistical weights, and partition functions, were taken from the HITRAN 2020
database \cite{Gordon2022}, except for C$_6$H$_6$ for which the
molecular parameters were provided based on the GEISA database
\cite{GEISA2020}. \BT{We provide further details about the C$_6$H$_6$ line list used in the next section.}

Protoplanetary disks are obviously not isothermal (vertically or
radially), but previous studies have shown that the LTE assumption is
a good first step approximation to determine the relative column
densities of molecules and physical parameters of the line-emitting
regions \cite{Carr2011,Salyk2011a}. Non-LTE effects can play a role
for the higher excited energy levels \cite{Meijerink2009} if the local
density is less than $\sim 10^{15}$ cm$^{-3}$. Differences up to factors
of 3 in inferred column densities have been found in LTE vs non-LTE
comparisons for the case of HCN ro-vibrational lines
\cite{Bruderer2015}. Since non-LTE effects are expected to be
comparable and in the same direction for different molecules, the
effect on column density ratios is expected to be smaller than such a
factor of 3.

Given the overwhelming presence of the C$_2$H$_2$ band in the MIRI
spectrum, we first fit the broad continuum bump (component I) between 12 and 17
$\mu$m using a $\chi^2$ approach (see Extended Data Figure 7, left) and
including the contribution of $^{13}$C$^{12}$CH$_2$ with a
C$_2$H$_2$/$^{13}$C$^{12}$CH$_2$ ratio of 35, half of the interstellar
medium value to account for two carbon atoms \cite{Woods2009}. In
order to avoid the contribution of the other molecular features, the
$\chi^2$ fit is computed using 4 spectral windows ($12.1-12.2 \ \mu$m,
$12.65-12.9 \ \mu$m, $14.6-14.85 \ \mu$m, and $15.5-15.7 \ \mu$m). We
then find a more optically thin model that reproduces well the main
$Q-$branches of C$_2$H$_2$ and $^{13}$C$^{12}$CH$_2$, assuming that
this second component II originates from a different region of the
disk. This model, with an emitting area \BT{corresponding to $R=0.07$~au}, is only
illustrative since the accuracy of this fit is limited by the lack of
spectroscopic data of $^{13}$C$^{12}$CH$_2$. \BT{Excited states of the $\nu_5$ band of $^{13}$C$^{12}$CH$_2$ are indeed missing in the molecular databases whereas the contribution of those states to the $Q-$branch of C$_2$H$_2$ is expected to be significant.}

The resulting C$_2$H$_2$ models for both component I and II are then subsequently used in the analysis of the other species to identify spectral windows that are free of contamination by narrow C$_2$H$_2$ features (see e.g., Extended Data Figure 8). However, in order to subtract the contribution of optically thick C$_2$H$_2$ (component I), we do not use the best fit model but subtract a spline fit through the two broad C$_2$H$_2$ bumps. This strategy avoids artefacts in the subtracted spectra that would be due to the imperfect C$_2$H$_2$ model. 

When subtracting the prominent C$_2$H$_2$ bumps prior to the fit of the other features, we implicitly neglect mutual line overlap between C$_2$H$_2$ and the other species which can be relevant if the emission of the species originates from the optically thick C$_2$H$_2$ component I. In fact, as discussed in the main text, only the analyses of CO$_2$ and HCN are affected by C$_2$H$_2$ opacity. \BTbis{This is because CO$_2$ and HCN features could originate from component I and are located close to prominent C$_2$H$_2$ lines. For HCN, masking by C$_2$H$_2$ is too significant to put constraints on the amount of HCN in component I. For CO$_2$, we conducted additional tests and find that the derived column densities of CO$_2$ for component I change by only a factor of less than two when including mutual shielding of the lines.}

The best fit slab model parameters ($N$, $T$, emitting area
characterized by $R$) for the species other than C$_2$H$_2$ are then estimated by a
$\chi^2$ approach (see Extended Data Figure 7). Extensive
grids of models varying the column density from $10^{15}$ up to
$10^{22}$ cm$^{-2}$, in steps of 1.26 in log10-space and temperature
from 100 up to 1500 K in steps of 25 K were computed. Given that
C$_2$H$_2$ emission is highly optically thick, the fit of the 13.7 $\mu$m
bump allows us to determine the emitting radius of $R = 0.033$
au. \BT{Interestingly, the fit of the CO$_2$ feature points toward a similarly small emitting
area. In contrast, for the C$_4$H$_2$ and benzene prominent features, a compact emission with $R\lesssim 0.07~$au is excluded. When the $\chi^2$ cannot constrain the emitting size (optically
thin lines), the same emitting radius of $0.07$~au is used to evaluate or place upper limits on column densities (CH$_4$,  C$_6$H$_6$, C$_4$H$_2$, HCN)}. \BTbis{For H$_2$O, the emitting area is unconstrained and we provide (upper limit) column densities for either component I or II in Table 1.}

In the calculation of the $\chi^2$, specific spectral windows are chosen to avoid contamination by other molecular features. The 1, 2, and 3 $\sigma$ confidence intervals are estimated by drawing the contours of $\Delta \chi^2_{red} = \chi^2_{red}(N,T)-\chi^2_{red,min}$ corresponding to values of \BTbis{2.3, 6.2, and 11.8}, respectively, and using a representative noise level of $\sigma=0.14~$mJy, where $\chi^2_{red}(N,T)$ is the reduced $\chi^2$ obtained by fitting the emitting area for a given value of $(N,T)$\cite[see e.g.,]{Avni1976}.




\medskip
\noindent \textbf{Benzene spectroscopy}

\BT{Benzene is included in the GEISA database\cite{GEISA2020}, but the existing line list does not provide Einstein A coefficients nor statistical weights and involves only the cold $\nu_4$ band centered at 14.837 $\mu$m  (673 .975 cm$^{-1}$). Therefore, the missing spectroscopic parameters that are necessary for the present study have been generated.}

\BT{For the cold $\nu_4$ band, we completed the line list, in terms of nuclear spin statistical weights and Einstein coefficients, using the method described in ref.\cite{Simeckova2006}, and the available spectroscopic constants\cite{Dang1989}. For the partition function, that involves a vibrational and a rotational contribution, we used the empirical equations of ref.\cite{Dang1989}. Before this, we made extensive calculations to check that these equations are usable for the  $50~$K $< T < 500~$K temperature range with an error that is less than 0.5\%. The contribution of hot bands at 14.9 $\mu$m is missing in the GEISA line list. For this heavy molecule, these hot bands contribute about ~45 $\%$ to the infrared activity at 14.9 $\mu$m  at room temperature. To account for these contributions, we generated empirical line lists, using the cold $\nu_4$ band as a “guide list” and the cross-section measurements of benzene performed at high resolution and for different temperatures\cite{Sung2016}.}

\medskip
\noindent \textbf{Details of disk models}

The observational results are compared with a number of
state-of-the-art thermochemical disk models in Extended Data Table 2. In this
comparison, we show two observational values of the ratios; assuming
either an emitting region corresponding to the highly optically thick
component I of C$_2$H$_2$, or from the less optically thick component II. The
thermochemical models assume a gas surface density structure, often
taken to be the self-similar solution of a viscously evolving disk.
Since we are only interested in the inner few au of the disk, the
precise shape and size of the outer disk are not relevant. The gas
distribution in the vertical direction is characterized by a scale
height and flaring index. The disk is irradiated by the star whose FUV
spectrum is given by its effective temperature $T_{\rm eff}$ and
strength by its luminosity. Extra UV due to accretion is modelled by
adding a $10^4$~K black body with a strength proportional to the
observed accretion rate or $L_{\rm FUV}$. X-rays and cosmic rays are
also included, the latter usually at a generic rate of $\sim 10^{-17}$
s$^{-1}$.

The models first solve for the dust temperature given the source's
luminosity, and then either assume that the gas temperature is equal
to the dust temperature or solve explicitly for the gas temperature by
iterating over the heating and cooling balance with a
small chemical network. Typical gas temperatures in the inner ($<1$
au) disk are a few thousand K at the top of the atmosphere, dropping to
several hundred K deeper in the disk where most of the molecular
emission originates \cite{Woitke2018}. Such temperatures are
consistent with our inferred values of 300--600 K for C$_2$H$_2$ and
other molecules. Finally, the 2D abundance distribution of each molecule can be
determined by solving the chemistry at each grid point using a more
extensive chemical network. Details can be found in refs.\cite{Najita2011,Bruderer2013,Woitke2018}.

Most relevant for comparison with our observations are those models
that include a large hydrocarbon network. In particular, Woods \&
Willacy \cite{Woods2009} have developed a detailed chemical model
appropriate for a disk around a T Tauri star including not just
$^{12}$C but also $^{13}$C isotopologues. We take the column densities
at 1 au from their Table~3 for comparison in Figure 3. Walsh et al.\
\cite{Walsh2015} have run chemical models for disks around a M dwarf, a
T Tauri star and a Herbig star, to investigate how the chemistry
differs across the stellar mass range. We focus here on their results
for the M-dwarf disk. Walsh et al.\ show not only total column
densities as function of disk radius, but also the column densities
above the dust $\tau=1$ surface at 14 $\mu$m, since mid-infrared
observations do not probe down to the midplane. We take the latter for
our comparison. None of these models vary the input volatile C and O
abundances (i.e., the amount of carbon and oxygen that can cycle
between gas and ice), which are usually taken such that oxygen is more
abundant than carbon at C/O=0.4.

Najita et al.\ \cite{Najita2011} have also presented sophisticated
inner disk models focusing on smaller molecules up to C$_2$H$_2$, HCN
and H$_2$O to investigate trends with disk parameters and C/O. Their
models include accretional heating and stellar X-rays, but 
not FUV radiation in the heating and chemistry. Their model abundances of
hydrocarbon molecules are clearly increased when C/O is increased. \BT{The same is found in more recent models by Woike et al. \cite{Woitke2018} and Anderson et al.\cite{Anderson2021}.}

\clearpage


\section{Extended Data}\label{secA1}
\renewcommand{\figurename}{Extended Data Figure}

\begin{figure}[htb!]
\centering
\includegraphics[width=15.5cm]{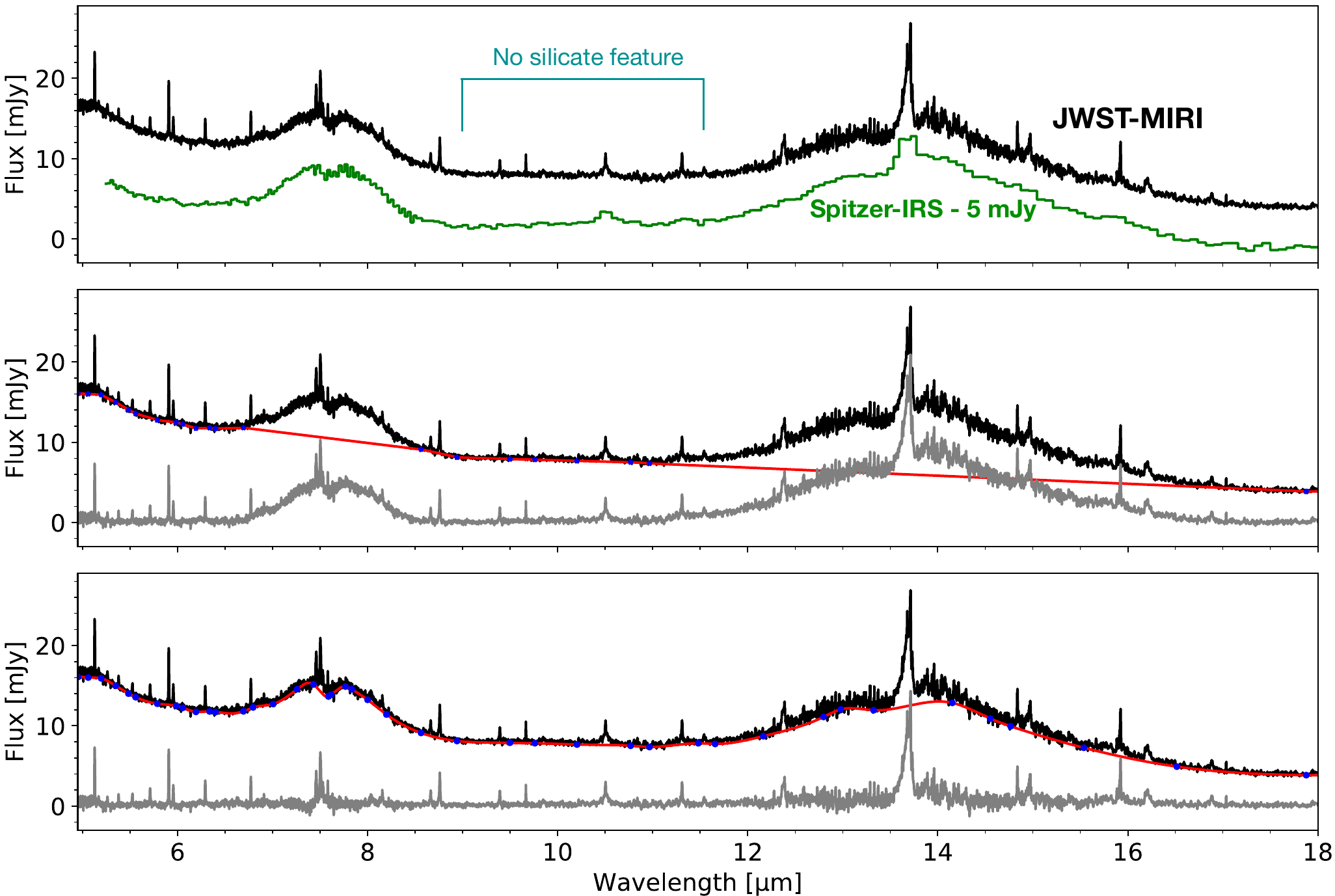}
\caption{\BTbis{Comparison between the \textit{Spitzer}-IRS spectrum and the MIRI-MRS spectrum and baseline fits of the MIRI-MRS spectrum. The \textit{Spitzer}-IRS low-resolution spectrum\cite{Lebouteiller2011} has been shifted by 5 mJy to ease the comparison with the MIRI-MRS spectrum (top panel). Baseline fits used in the continuum subtracted spectrum presented in Figure 1 and 2 in the Results section are shown in the middle and bottom panel, respectively.} The blue dots represent the location where the continuum is evaluated. The red curve is the interpolated continuum used to produce continuum-subtracted spectra (in grey). The presence of warm dust is evidenced by the infrared continuum emission on either sides of the two C$_2$H$_2$ bumps but no silicate feature is detected. HI and H$_2$ lines are present in the spectrum and will be analysed in a next paper.}
\label{FigA1}
\end{figure}

\begin{figure}[htb!]
\centering
\includegraphics[width=\linewidth]{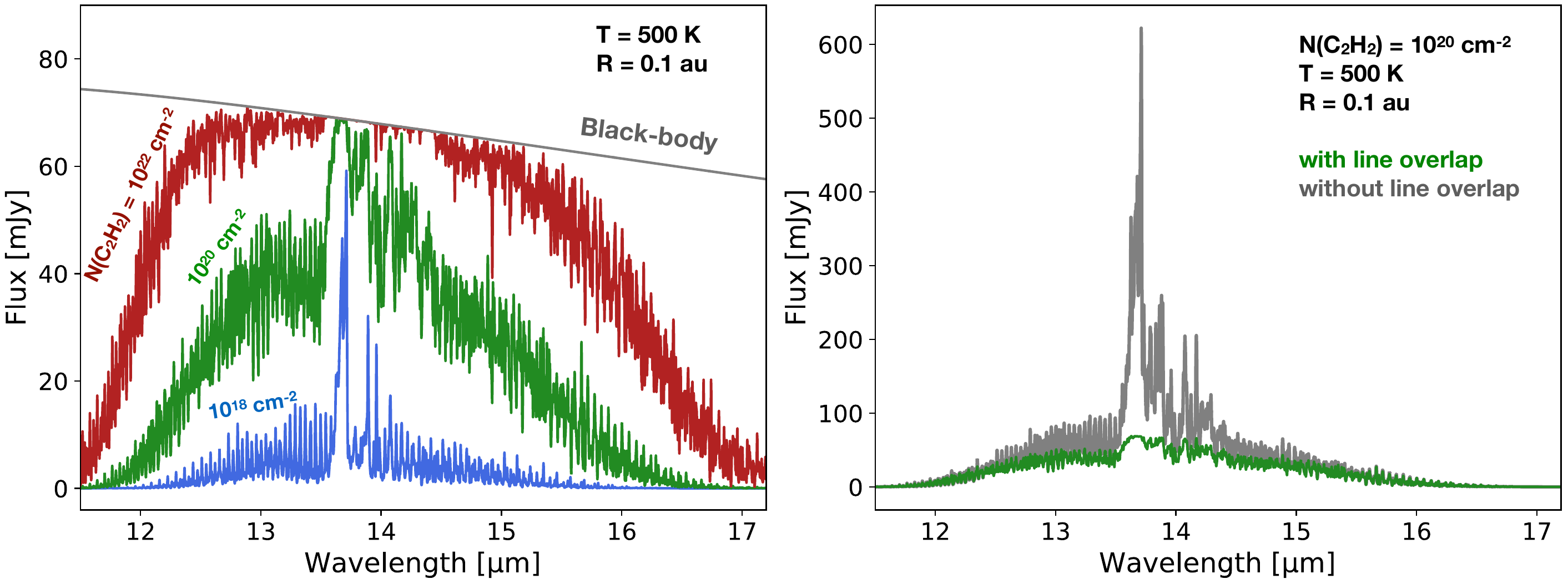}
\caption{\BTbis{Effect of line overlap on the main C$_2$H$_2$ feature at 13.7 $\mu$m.} \textit{Left:} C$_2$H$_2$ emission as function of column density for $T=500$~K and $R=0.1~$au. Note that the $Q$-branch becomes highly optically thick above $N($C$_2$H$_2)=10^{20}$ cm$^{-2}$ and flattens. The contrast between the amplitude of the narrow features on either side of the $Q-$branch and the continuum level decreases by increasing $N($C$_2$H$_2)$. A column density of at least $N($C$_2$H$_2) \simeq 10^{20}$ cm$^{-2}$ is required to fit the observations. \textit{Right:} Importance of line overlap in slab models. For highly optically thick lines that are close to each other such as in the $Q-$branch of C$_2$H$_2$, slab models neglecting line-overlap overestimate the fluxes. For C$_2$H$_2$, this effect dominates for $N \gtrsim 10^{19}$
  cm$^{-2}$.}
\label{FigA2}
\end{figure}

\begin{figure}[htb!]
\centering
\includegraphics[width=\textwidth]{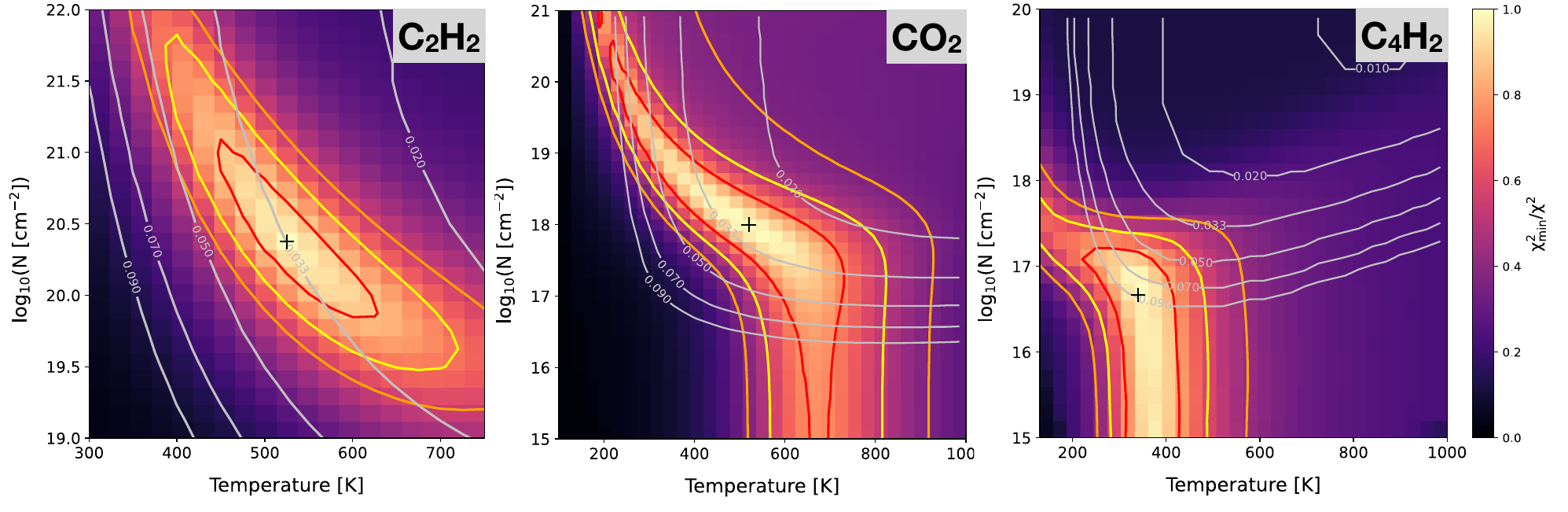}
\caption{\BTbis{Constraints obtained from $\chi^2$ fits.} The $\chi^2$ maps for the fit of the 13.7 $\mu$m broad bump associated with C$_2$H$_2$ (left), and the CO$_2$ (middle) and C$_4$H$_2$ (right) features are shown. The $1 \sigma$, $2 \sigma$, and  $3 \sigma$ confidence intervals are pictured in red, yellow, and orange, respectively. The best-fitting emitting radius $R$ for each value of $N$ and $T$ is indicated as grey lines. In general, we find a degeneracy between a high $T$ and low $N$ solutions, and a low $T$ and high $N$ solutions. For CO$_2$ the best fit corresponding to an emitting area of 0.033~au is chosen to alleviate the degeneracy and compare with the optically thick component of C$_2$H$_2$ (component I). We note that for $R=0.07$~au, corresponding to component II, the CO$_2$ feature can be fitted by either a hot and thin model or a cold and thick model. However, the thick solution over-predicts $^{13}$CO$_2$ emission which is not detected. We therefore report in Table 1 the column density of the optically thin solution for component II.} 
\label{FigA3}
\end{figure}

\begin{figure}[htb!]
\centering
\includegraphics[width=14cm]{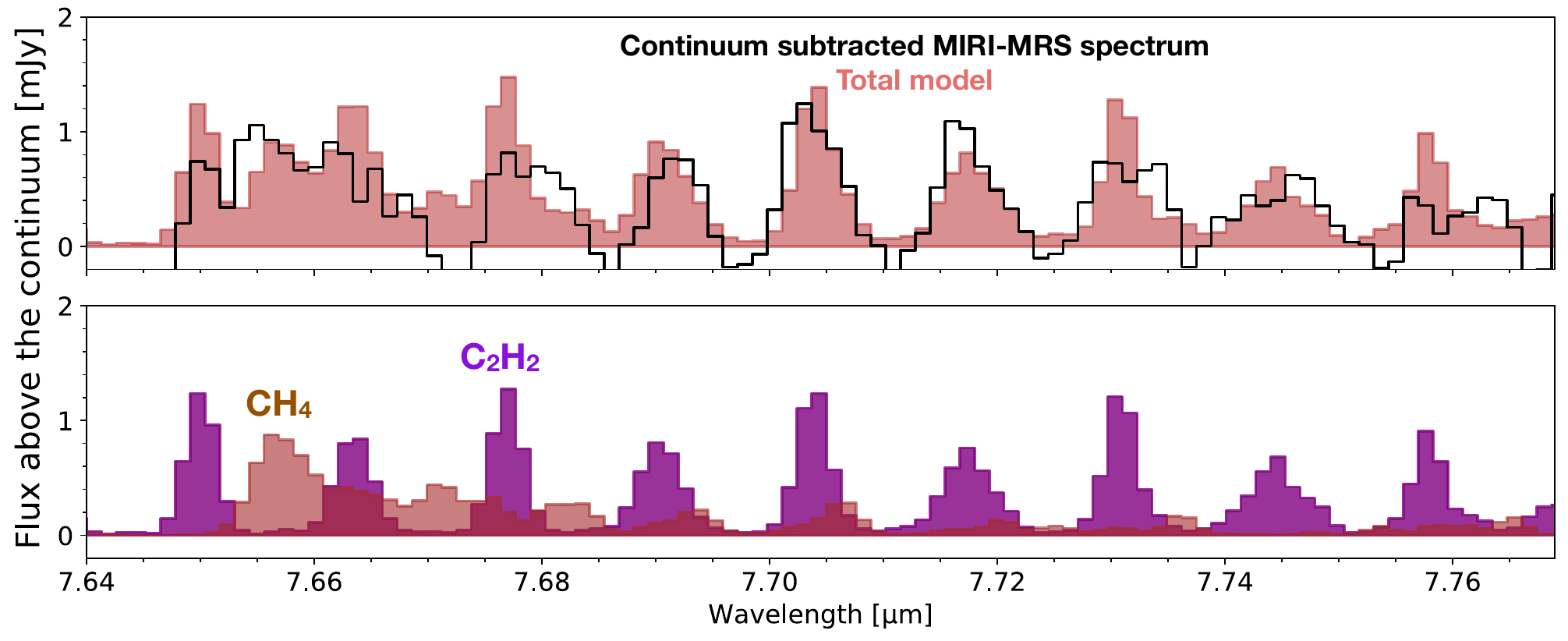}
\caption{ \BTbis{Possible indication for CH$_4$ emission in the 7.64-7.77 $\mu$m range. CH$_4$ emission could be present at 7.655 $\mu$m in addition to the many C$_2$H$_2$ lines in this region. The column density of CH$_4$ is estimated assuming that the emission originates from component II (see main text, Table 1)}. The C$_2$H$_2$ model in purple corresponds to the component II \BT{for which the best-fit column density has been increased by a factor of 4 to better match the series of C$_2$H$_2$ lines in that specific spectral region.}}
\label{FigA4}
\end{figure}

\begin{figure}[htb!]
\centering
\includegraphics[width=15cm]{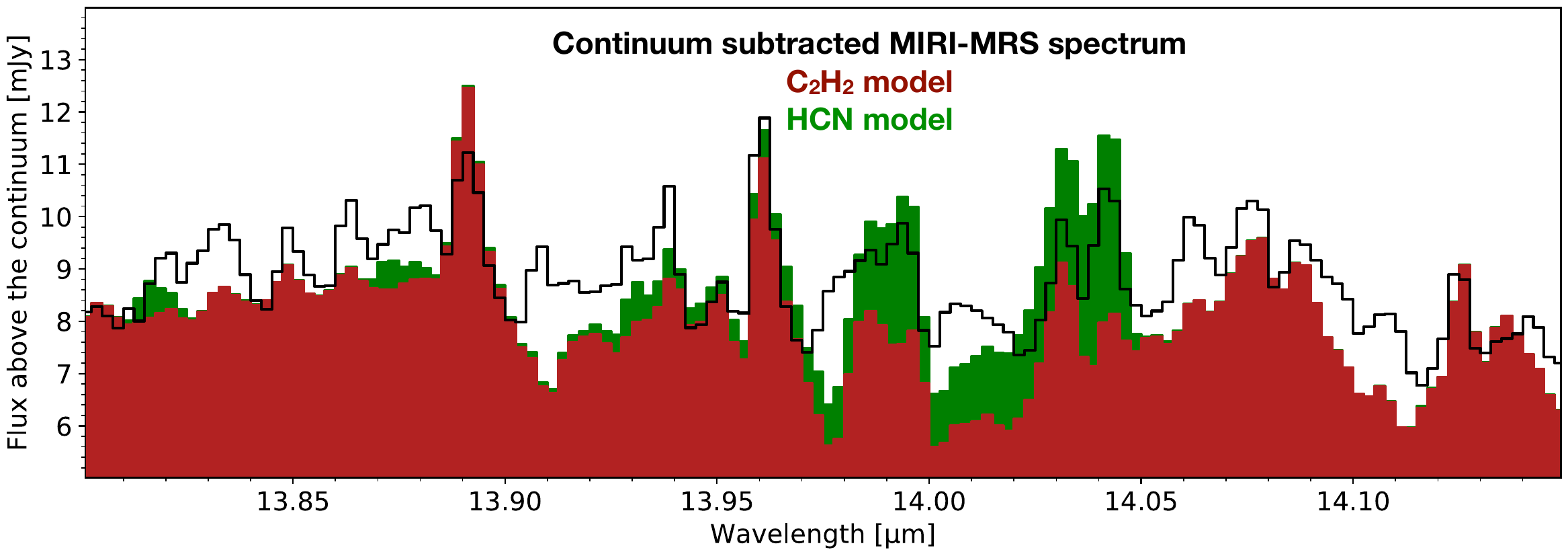}
\caption{\BTbis{Constraints of the amount of HCN in the 14~$\mu$m region.} The C$_2$H$_2$ model, including both \BTbis{component I and II} is shown in red on top of the MIRI spectrum where the contribution of the C$_2$H$_2$ thick component is not subtracted. This figure shows that a maximum column density of HCN of $N= 1.5\times 10^{17}~$cm$^{-2}$ can be hidden in the C$_2$H$_2$ line forest in this region \BT{assuming an origin in the optically thin component II ($R=0.07$~au and $T=400~$K). HCN emission from the C$_2$H$_2$ thick component I would be highly masked by C$_2$H$_2$ and therefore its column density remains unconstrained.}}
\label{FigA5}
\end{figure}

\begin{figure}[htb!]
\centering
\includegraphics[width=\linewidth]{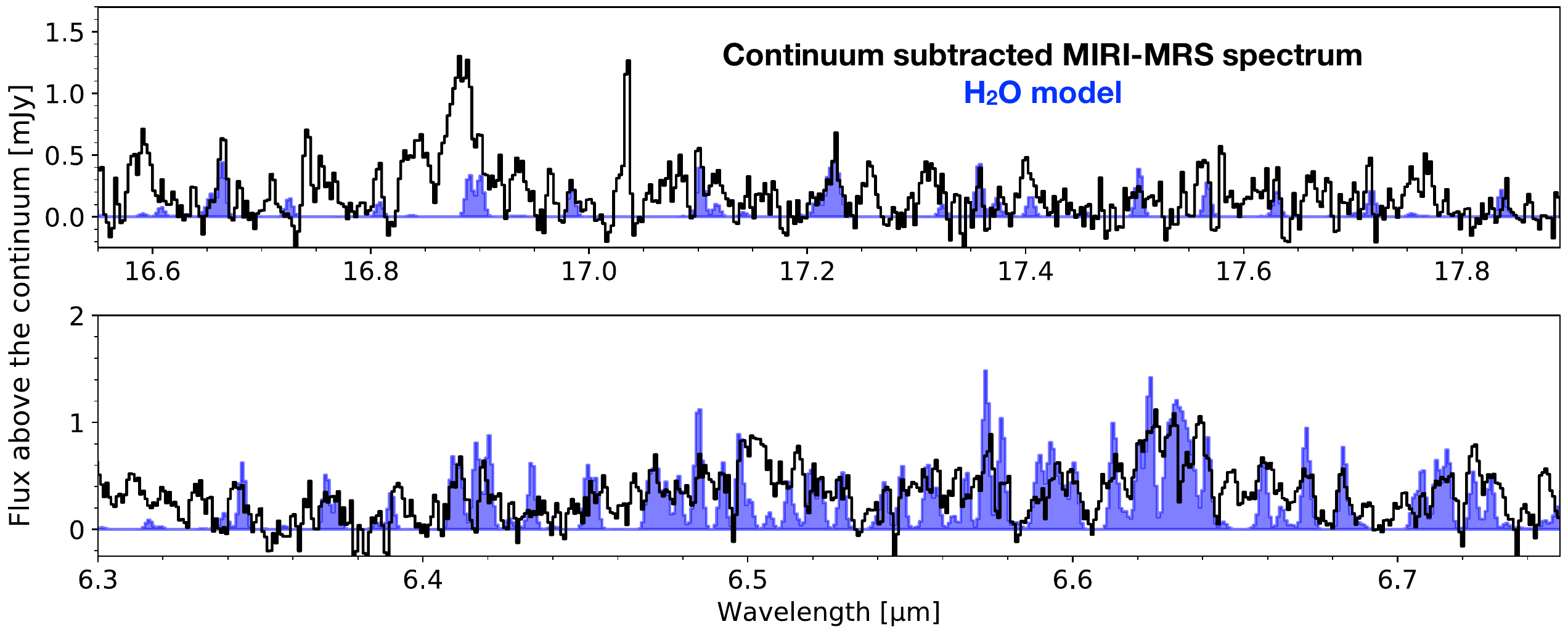}
\caption{Possible detection of weak H$_2$O
  lines in the 17.2 $\mu$m and 6.5 $\mu$m regions. The pure rotational lines at
  16.5--18 $\mu$m can hide as much as $N$(H$_2$O)=$3 \times 10^{18}$
  cm$^{-2}$ assuming a fixed temperature of 525 K and a characteristic emitting radius of $R=0.033~$au, corresponding to the optically thick C$_2$H$_2$ \BTbis{component I}. These lines are not affected by masking of C$_2$H$_2$ since only very weak lines of C$_2$H$_2$ are present in these spectral ranges. Some lines in the 6.3-6.8 $\mu$m range are somewhat overestimated by our LTE model but non-LTE effects will tend to quench these lines compared to the pure rotational lines longward of $\sim 12 \mu$m \cite{Banzatti_2023_H2O}.}
\label{FigA6}
\end{figure}

\begin{figure}[htb!]
\centering
\includegraphics[width=8cm]{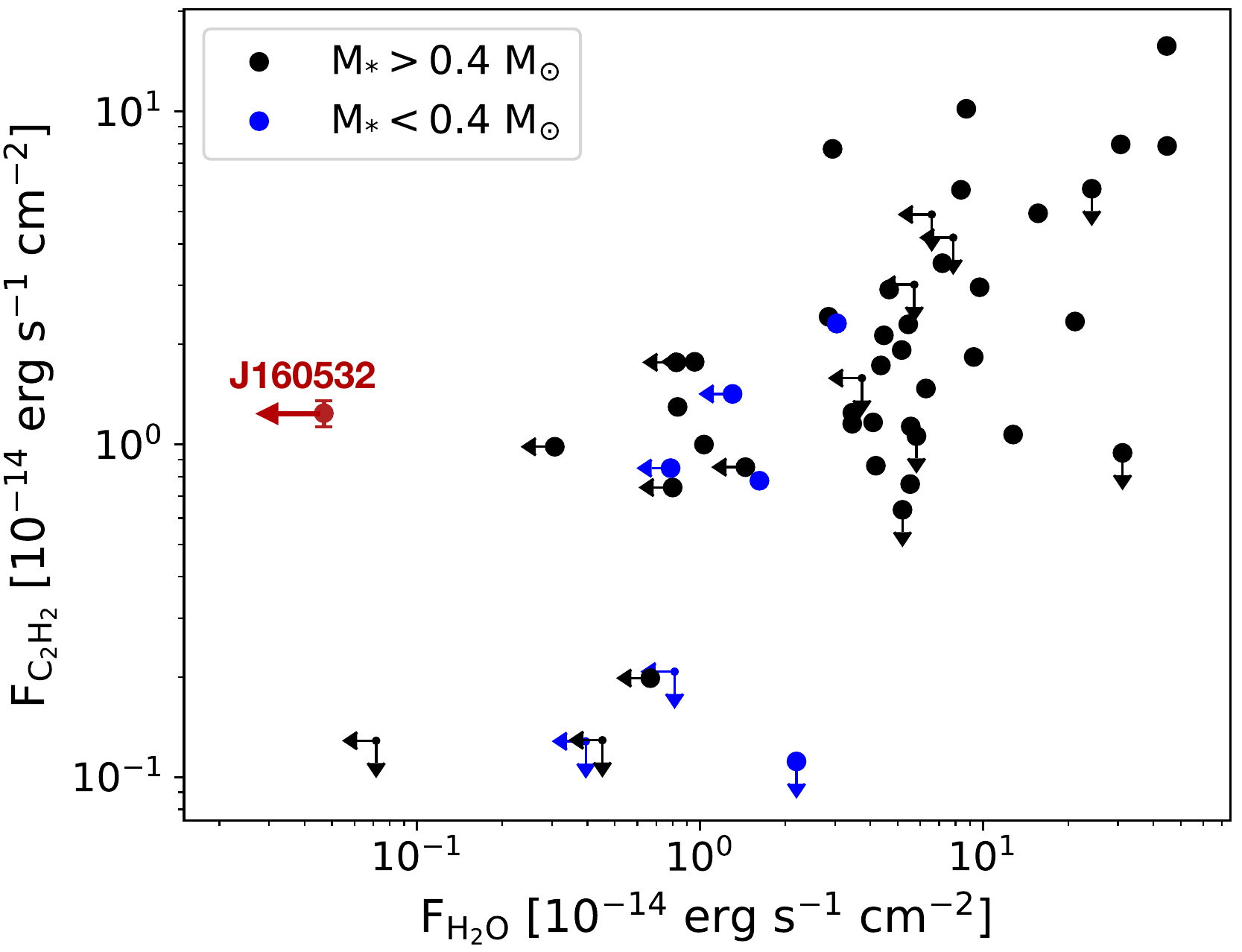}
\caption{\BTbis{J160532 line fluxes compared to other disks. This figure presents} a comparison of C$_2$H$_2$ versus H$_2$O flux scaled to 140 pc for a number of T Tauri disks observed with {\it Spitzer}
\BT{and complied by ref.}\cite{Banzatti2020} and J160532 observed with JWST MIRI. \BT{The line fluxes for J160532 are consistently calculated by integrating the flux of the three water features at 17.12 $\mu$m, 17.22 $\mu$m, and 17.36 $\mu$m, and the C$_2$H$_2$ feature over a window between 13.65-13.72 $\mu$m as explained in ref.\cite{Banzatti2020}. Leftward (resp. downward) arrows represent upper limits on H$_2$O (resp. C$_2$H$_2$) line flux.}}
\label{FigA7}
\end{figure}

\renewcommand{\tablename}{Supplementary Table} 

\begin{table}[h]
\begin{center}
  \caption{Observed column density ratios of various species compared with disk models \BT{with solar C/O elemental ratio in the gas-phase}.}
\begin{tabular}{l|cc|cc}
  \toprule
  Ratio & \multicolumn{2}{c|}{Observations$^a$}  & TTauri model & BD model  \\[0.5ex]
        &   Component I  &   Component II & WW 09$^2$ & WNvD15$^3$ \\
  \midrule
 C$_4$H$_2$/C$_2$H$_2$ &  - & 0.3 &    0.18 &    -  \\
 CH$_4$/C$_2$H$_2$     &  - & 0.6 &    0.18 &    -  \\
 C$_6$H$_6$/C$_2$H$_2$ & -  & 0.3   &   0.02$^4$ & - \\
 CO$_2$/C$_2$H$_2$     &  $8~10^{-3}$ & 0.14  &  2.0  &  -   \\
 H$_2$O/C$_2$H$_2$     & $\le 1~10^{-2}$   &  $\le$3 &   192 & 100 \\
  \bottomrule
\end{tabular}
\end{center}
{
1. Assuming that the emission is either confined to the C$_2$H$_2$ highly optically thick \BTbis{component I} or the optically thinner \BTbis{component II}. \BT{The prominent features of the small hydrocarbons are associated to the extended component II only whereas the CO$_2$ emission originates more likely from the thick component I (see main text).} 2. Woods \& Willacy (2009). 3. Walsh, Nomura \& van Dishoeck (2015). 4. Woods \& Willacy (2007).}
\label{Tab2}
\end{table}

\clearpage

\section*{Data Availability}

\BTbis{The original data analysed in this work are part of the GTO-MIRI programme "MIRI EC Protoplanetary and Debris Disks Survey" (ID 1282) with number 47 and will become public on August 1 2023 on the MAST database \url{https://archive.stsci.edu/}.
The continuum-subtracted spectra presented in Figure 1 (right) and in Figure 2 will be made available on Zenodo upon publication.
The spectroscopic data for all the species but benzene are available on the HITRAN database (\url{https://hitran.org/}). For benzene, the data will be shared on  request
to the corresponding author.} \\

\section*{Code Availability}

\BTbis{The slab model used in this work is a private code developed by BT and collaborators. It is available from the corresponding author upon request. The synthetic spectra presented in this work can be reproduced using the \texttt{slabspec} code, which is publicly available at \url{https://doi.org/10.5281/zenodo.4037306}.} \\

\section*{Acknowledgements}
The MINDS team would like to thank the entire MIRI European and US
instrument team. Support from StScI is also appreciated. The following
National and International Funding Agencies funded and supported the
MIRI development: NASA; ESA; Belgian Science Policy Office (BELSPO);
Centre Nationale d’Etudes Spatiales (CNES); Danish National Space
Centre; Deutsches Zentrum fur Luftund Raumfahrt (DLR); Enterprise
Ireland; Ministerio De Economi{\'a} y Competividad; Netherlands
Research School for Astronomy (NOVA); Netherlands Organisation for
Scientific Research (NWO); Science and Technology Facilities Council;
Swiss Space Office; Swedish National Space Agency; and UK Space
Agency.

B.T. is a Laureate of the Paris Region fellowship
program, which is supported by the Ile-de-France Region and has
received funding under Marie Sklodowska-Curie grant agreement
No. 945298. G.B. thanks the Deutsche Forschungsgemeinschaft (DFG) - grant 325594231,
FOR 2634/2. E.v.D. acknowledges support from the EU ERC grant 101019751
MOLDISK and the Danish National Research Foundation through the Center
of Excellence “InterCat” (DNRF150). D.G. would like to thank the
Research Foundation Flanders for co-financing the present research
(grant number V435622N). T.H. and K.S. acknowledge support from the
ERC Advanced Grant Origins 83 24 28. I.K., A.A., and E.v.D.
acknowledge support from grant TOP-1614.001.751 from the Dutch
Research Council (NWO). I.K. and J.K. acknowledge funding from
H2020-MSCA-ITN- 2019, grant no. 860470 (CHAMELEON). O.A. and
V.C. acknowledge funding from the Belgian F.R.S.-FNRS. I.A. and
D.G. thank the European Space Agency (ESA) and the Belgian Federal
Science Policy Office (BELSPO) for their support in the framework of
the PRODEX Programme. D.B. has been funded by Spanish
MCIN/AEI/10.13039/501100011033 grants PID2019- 107061GB-C61 and
No. MDM-2017-0737. A.C.G. has been supported by PRIN-INAF MAIN-STREAM
2017 and from PRIN-INAF 2019 (STRADE). T.P.R acknowledges support from
ERC grant 743029 EASY. D.R.L.  acknowledges support from Science
Foundation Ireland, grant number 21/PATH-S/9339. L.C. acknowledges
support by grant PIB2021-127718NB-I00, from the Spanish Ministry of
Science and Innovation/State Agency of Research
MCIN/AEI/10.13039/501100011033.

\section*{Author contributions} 
BT and GB did the analysis using molecular data files created by
AA and AP and a model developed by BT and JB. GB, SG and DG performed the data reduction, supported by IA,
JS, MS, GP, VC, and JB.  EvD, BT, and GB wrote the manuscript. TH and IK
planned and co-led the MIRI guaranteed time project on disks. All
authors participated in either the development and testing of the MIRI
instrument and its data reduction, in the discussion of the results,
and/or commented on the manuscript.

\section*{Competing Interests}
The authors declare no competing financial interests.\\

\section*{Correspondence} Correspondence and requests for materials
should be addressed to Benoit Tabone,
benoit.tabone@universite-paris-saclay.fr.






\clearpage

\bibliography{biblio.bib}




\end{document}